# DFT mediated $X_2AuYZ_6$ (X= Cs, Rb; Z= Cl, Br, I) double Perovskites for photovoltaic and wasted heat management device applications

S. Mahmud[1,2], M.A. Ali[1,*], M. M. Hossain[1], M. M. Uddin[1]

[1]Advanced Computational Materials Research Laboratory, Department of Physics, Chittagong University of Engineering and Technology (CUET), Chattogram-4349, Bangladesh.

[2]Department of Electrical and Electronic Engineering, Jatiya Kabi Kazi Nazrul Islam University (JKKNIU), Mymensingh-2224, Bangladesh.

*Abstract*

This paper presents the phase stability, opto-electronic and thermo-electric behavior of $X_2AuYZ_6$ (X = Cs, Rb; Z = Cl/Br/I) double perovskite halides by using the DFT method. The compounds belong to the cubic arrangement and are verified by the tolerance and octahedral factor. Formation enthalpy and binding energy meet the requirements of structural stability. The ductility behavior was also confirmed by the Cauchy pressure, Pugh's ratio, and Poisson's ratio. The positive frequency of phonon dispersion except $Rb_2AuYI_6$ compound shows the dynamical stability and the negative formation energy of each identified competing phase confirms the thermo-dynamic equilibrium of all compounds. The band gap values of 2.85(2.91), 2.35(2.40), and 1.74(1.78) eV of $Cs_2AuYZ_6$ ($Rb_2AuYZ_6$) [Z = Cl, Br, I) double perovskites has been explored in the context of optoelectronic properties, and the results show that these materials might be useful in such devices. The spectral optical response covers the visible-to-UV area, which governs the solar cell and thermo-electric device applications. A comprehensive study of thermo-electric properties such as the thermal conductivity (electrical and electronic part), carrier concentration, thermo-power, and figure of merit was also observed. The investigated compounds [Cs (Rb)-based] exhibit ZT values of 0.51(0.55), 0.53(0.62), and 0.58(0.75) at room temperature with Cl, Br, and I respectively. Additional routine work was also done on the thermo-mechanical characteristics. These studies provide in-depth knowledge of these materials in preparation for their future use.

Keywords: DFT, Halide double perovskites, Stability, Opto-electronic properties, Figure of merit, SLME, thermo-dynamic properties.

## 1. Introduction

The global energy problem is becoming worse due to consumption every day. The combustion of fossil fuels, such as coal, oil (petroleum), and natural gas, has contributed to environmental problems like air pollution and climate change by releasing carbon dioxide and other pollutants into the atmosphere for many years. To lessen these negative effects on the environment, attempts are being made to switch to greener and more sustainable energy sources. The two alternative main sources of energy are sunlight and waste heat, which may be used in photovoltaic and thermo-electric processes. Material choice affects the capacity to carry out both thermal and photoelectric applications. The development of novel solar cell materials and technologies with significant efficiency, strong inherent stability, and less costly manufacturing is of major importance to the industry. The perovskite halide materials have great interest in the above processes. Perovskite halides, which offer a flexible platform for a variety of applications, with solar energy at the forefront, constitute a paradigm shift in the field of materials research. The general structure of such perovskite is represented as $ABX_3$ or $A_2BX_6$, called single or double perovskite, respectively, with the A cation positioned at the corner, the B cation at the centre, and the X anion at the face centre. A single perovskite can only tolerate the $2^+$ cationic arrangement in B-sites, whereas a double perovskite can manage a wide range, from 1 to $4^+$[1]. This is why the double perovskite offers improved functionalities, tunabilty properties, enhanced stability or long term application, reduced defect density, and flexibility in band gap engineering.

The PCE (power conversion efficiency) of oxide-based perovskites has increased by about 21%, indicating their increased efficiency for solar applications[2]. The hybrid halide, or organic-inorganic lead halide, is the most studied material for the advancement of solar cell technology due to its high spectroscopic limited maximum efficiency (SLME) of 3 to 22% from 2009 to till date[3–6]. However, toxicity and instability are the two fundamental problems with lead-based halide compounds. The toxic element of the $Pb^{2+}$ ion is substituted by monovalent along with trivalent elements in order to balance the B-site and create the double perovskite structure $A_2^{+1}B^{+1}B^{+3}X_6$, which solves the problems. It has been discovered that lead-free double perovskites exhibit higher solar efficiency (till 30.95%) and are safe for use while being more stable at appropriate temperatures and humidity levels than single perovskites[7].

Extensive research has been carried out on inorganic lead-free double perovskite materials having the chemical formula Cs/Rb$_2$B$^+$B$^{3+}$X$_6$, where B$^+$ represents Ag+, Au+, Tl+, and B$^{3+}$ represents Bi$^{3+}$, Sb$^{3+}$, In$^{3+}$, and X represents I$^-$, Br$^-$, Cl$^-$ [8–12]. Such substances have a lot of potential to improve technological advances in many different fields, may be producing environmental friendly and productive gadgets. For example, E. T. McClure *et al*. conducted a study on Cs$_2$AgBiCl$_6$ and Cs$_2$AgBiBr$_6$, which had indirect band gaps of 2.77 eV and 1.95 eV, respectively[13]. These band gaps surpass the Shockley–Queisser band gap limit for the optimal performance of solar cells, leading to efficiency reductions of up to 2.5%. A.J. Kale gave a presentation on the DFT analysis of CS$_2$AuBiCl$_6$ for single junction solar cells, which has a band gap of 1.40 eV and a PCE of 22.18%[14]. A U Haq *et al*. examined the optoelectronic and thermal behavior of the Rb$_2$XGaBr$_6$ (X= K, Na) compound, revealing electronic band gap values of 1.90 and 2.2 eV, respectively[15]. S.A. Aldaghfag *et al*. determined the direct band gap values of 3.65 and 3.63 eV for K$_2$ScAgCl$_6$ and Na$_2$ScAgCl$_6$, respectively, as sunlight absorbing materials and UV photo detectors[16]. Recently, lead-free perovskites of Cs$_2$AgBiX$_6$ (X = Cl, Br, and I) and Cs$_2$AgSbX$_6$ (F, Cl, Br and I) have gained appeal due to their capacity to handle some of the most significant issues such as band gap tuning and toxicity[17–23]. Some perovskites exhibit excellent stability in the air, but their outsized band gap values (over 2.2 eV) reduce their photovoltaic capacity[24-25]. The results on the PCE of Au-based DP, Cs$_2$AuBiCl$_6$, [14] influence us for searching new Au-based DPs with a band gap of around Shockley–Queisser (S.Q) limit, from which a significant PCE is expected.

A theoretical limit on the highest efficiency of a solar cell is known as the spectroscopic limited maximum efficiency (SLME), which is based on the idea of detailed balancing (S.Q limit)[26] and takes into consideration the whole solar spectrum. This study examines the basic physical mechanisms that control the transformation of photons into electrical energy in a solar cell. Unlike the S.Q limit, which provides the theoretical highest efficiency based only on semiconducting band gap (i.e., extended S.Q limit) with recombination losses. The SLME offers a more thorough assessment of the highest possible efficiency for a solar cell by taking into account a wider variety of physical processes.

Besides PV, the researcher have studied the Pb free DP materials for thermo electric application by using wasted heat for green electricity generation. When a thermoelectric material achieves a

ZT value of 1.0, it is considered to be operating at an exceptionally high efficiency when it comes to turning heat into electricity. Remarkably, halide DP materials have mostly been explored for photonic applications, with less experimental attention on their thermoelectric efficiency, despite their poor heat conductivity owing to cation structure and high charge mobility. Nonetheless, research into the thermoelectric capabilities of halide perovskites is becoming more and more popular. Several DPs, including $Cs_2AgBiX_6$ (X= Cl, Br)[27], $Cs_2ScTlX_6$ (X= Cl, Br, I)[28], and $Cs_2AuScI_6$[29], exhibited remarkable figure of merit values for thermoelectric applications at room temperature, which were approximately 0.75, 0.60, and 0.62, respectively. These prospective results encourage us to search for new potential DPs with a high ZT value.

We selected the six halide compounds by the following element: An alkali metal cations of Cs and Rb that can influence the stability and structure of compound. Gold (Au) has exceptional electrical conductivity, characterized by its high conductivity and low resistance. The electronic transport capabilities of DP may be influenced by this characteristic. Yttrium (Y) is a stable transition metal, which helps to improve the overall stability and longevity of materials, especially under challenging situations. Halides (Cl, Br and I) may cause modifications in the band structure, binding energies, and electronic transitions, which in turn affect the material's optical response.

It has been demonstrated that research on $X_2AuYZ_6$ (X = Cs, Rb, Z = Cl, Br and I) covers the Uv to visible band gap range of 2.91 to 1.74 eV, while $X_2AuYI_6$ (A= Cs, Rb) has a narrower band gap value of less than 2.2 eV, potentially making it suitable for photovoltaic purposes in renewable energy devices. The cationic combination of Y and Au at the B site offers two main advantages: achieving a broad to narrow band gap and reducing toxicity. Thus, the goal of this work is to use density functional theory (DFT) for finding the spectroscopic limited maximum efficiency by conducting a thorough and in-depth assessment of the thermodynamic phase stability, electronic, transport, optical and thermo-mechanical properties of aforementioned double perovskite halides.

Therefore, in this paper, $X_2AuYZ_6$ has the requisite stability, a satisfactory band gap, notable absorbance and efficiency, making it a compelling contender (especially I-based compounds) for green energy harvesting based on our calculations.

## 2. Computational technique

The titled task has been performed by the Wien2k software, which is based on the FP-LAPW (full potential linearized augmented plane wave) method by using electronic wave function [30]. The LAPW approach separates the unit cell of materials into two regions: the interstitial area, which uses plane waves, and the atomic sphere (also known as the muffin tin), which expands wave functions. Firstly, the structural optimization (E-V) was calculated by the GGA-PBE (Generalized Gradient Approximation-Perdew Burke Ernzerhof) [31] approach by implementing the 12×12×12 k point. For this, the initializing parameters were set as: RMT*$K_{max}$ = 8, $G_{max}$ = 14, charge convergence = $10^{-5}$ Ry, and energy convergence = 0.0001e. After optimization, we get the structural entities such as the lattice parameter, bulk modulus, minimum volume and energy by fitting the Vinet Rose equation of state [32]. Secondly, the TB-mBJ (Trans Blaha modified Becke-Johnson) approach was used to get a more reliable opto-electronic band gap that is close to experimental and time-efficient[33]. The spin orbit coupling method (TB-mBJ+SOC) and HSE-06 functional were used for accurate clarification of band gap measurement. For photovoltaic efficiency measurement, the SLME model was used by using the absorbing coefficient and thickness. The thermo-electric properties were calculated using Boltzmann transport theory and Boltztrap2 code [34]. The study of thermo-elastic characteristics was conducted using the IR-elast package [35]. Furthermore, the finite displacement method (FDM) was used to compute the phonon dispersion and associated partial DOS.

## 3. Results and discussion

### 3.1 Structural entities and stability criteria

**Fig. 1** shows the atomic 3D arrangement of the cubic crystal unit cell of Cs and Rb based halides with Cl, Br and I. The unit cells are face-centered with the Fm-3m (#225) space group. The corner-sharing element of Cs/Rb and the other octahedral elements of Au and Y occupy the positions of 8c (1/4, 1/4, 1/4), 4a (1/2, 1/2, 1/2), and 4b (0, 0, 0), while halide (X) ions are positioned at 24e (1/4, 0, 0), respectively. Firstly, the entity of ground state structure was estimated using the GGA-PBE approximation using volume optimization method which is presented in **Table 1**. For this, the Vinet-Rose equation of state [32] is employed for fitting the unit cell energy vs volume data and displayed in Fig. S1 (supplementary). It is noted that the lattice parameter increases from 10.7811 (10.7261) to 11.3529 (11.2605) Å and 12.1495 (12.0988) Å as the halide (X) ionic radius increases for Cs$_2$AuYX$_6$ (Rb$_2$AuYX$_6$), which dictates a decrease (less negative) in binding energy. As the lattice parameter and bulk modulus are inversely connected, the bulk modulus decreases from 32.10(33.25) to 27.35(27.64) GPa and 22.19(22.75) GPa of same series of halides. Furthermore, the compounds have four formula units in the ratio of 2:1:1:6.

Then, we begin by assessing the materials' structural stability using several stability index factors. Generally, double perovskites (DP) with the usual equilibrium A$_2$BB'X$_6$ follow a typical cubic crystal shape. However, this perfect structure might not always be the case because of things like size or dimension effects, distortion of octahedral geometry, changes of X and J-T (Jahn–Teller) effects[36-37].

The key parameter of the cubic sign or structural stability compound is ensured by its tolerance ($T_F$) and octahedral factor (µ), which calculated by the following equations of Goldsmith [29]:

$$T_F = \frac{R_A + R_X}{\sqrt{2}(\frac{R_{B'} + R_{B''}}{2} + R_X)}$$

$$\mu = \frac{R_{B'} + R_{B''}}{2R_X}$$

Where R denotes the RD Shannon ionic radius of the corresponding elements, A = Cs$^+$/Rb$^+$ with coordination number 12, $R_{B'}$ = Au$^+$, $R_{B''}$ = Y$^{3+}$, and R$_x$ = Cl$^-$/Br$^-$/I$^-$ with coordination number 6 [38]. Goldsmith finds that the ranges for tolerance and octahedral factor are 0.80 to 1.1 [39] and 0.414 < µ,< 0.89 [40] for structural stability.

The difference between total energy and the added energy of each element per atom measures the formation energy, and the difference between total energy and the added free state energy of each element measures the binding energy. The formation and binding energies of negative values confirm the stability issues, and more negative values indicate better stability. The following equations were considered for the calculation of formation and binding energy by the element [41]:

$$\Delta H_f = [E_{Cs_2AuYX_6} - (aE_{Cs} + bE_{Au} + cE_Y + dE_X)]/N$$

$$\Delta H_b = [E_{Cs_2AuYX_6} - (a\mu_{Cs} + b\mu_{Au} + c\mu_Y + d\mu_X)]/N$$

Where the compounds total energy with X= Cl, Br or I atoms is represented by $E_{Cs_2AuYX_6}$. The amount of energy of each atom is shown by the symbols $E_{Cs}, E_{Au}, E_Y$ and $E_X$ while a, b, c, and d refer for the number of atoms of Cs, Au, Y, and Cl/Br/I, in the unit cell or compositions, respectively. The formation energy increases from -2.45 (-2.41) to -2.07 (-2.02) eV and -1.62 (-1.57) eV with Cl, Br, and I atom compositions of Cs (Rb), indicating a decreasing trend in stability. Also, the binding energy also increases from -4.18 (4.16) to -3.81 (3.78) eV and -3.41 (3.37) eV due to the increasing halide ion substitution for same atom, respectively.

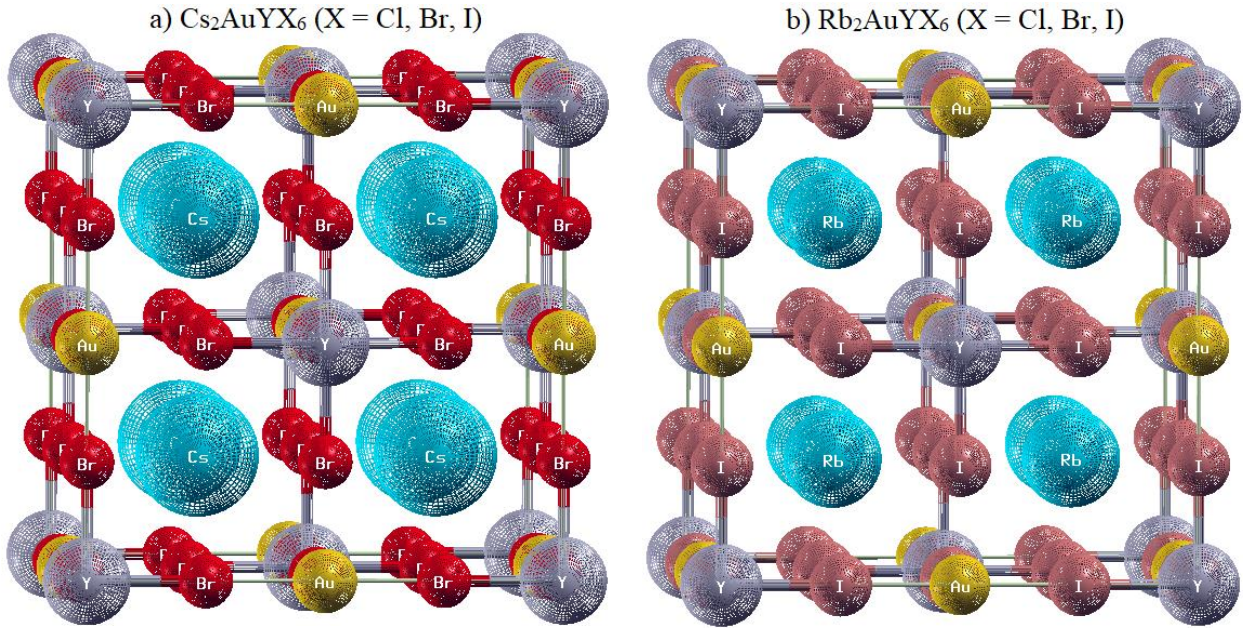

**Fig.1**: Unit cell of a) $Cs_2AuYX_6$; and b) $Rb_2AuYX_6$ [3D structure]

The dynamic equilibrium was checked by the phonon band structure or dispersion along with the phonon DOS (TDOS and PDOS). **Figure 2(a-f)** displays a well calibrated dispersion curve with PDOS. It describes how phonons in a crystalline material relate to one another in terms of frequency (energy) and wave vector (momentum) in the first Brillion zone. In a crystal lattice, collective atomic vibrations are represented by quanta of vibrational energy called phonons. Some important physical processes that phonons affect are heat transmission, electrical conductivity, and the thermal growth of objects. The dispersion showed a positive frequency except $Rb_2AuYI_6$ with a 3N vibrational mode. Among them, three are called the acoustic mode (green color), which operates in a low frequency region to converge gamma point only, and the remaining are called the optic mode and operate in a higher frequency region (Black color). From the partial DOS of phonon, we concluded that the band structure of phonon is dominated by the halide element in concert with slightly Au atom in the higher part of frequency range but in the lower part of frequency range (0-1 THz), they are consisted mostly by the Cs atom.

There is a larger phononic gap or phase space in the I-based compound than in the others, which might be the result of phonon heat being restricted due to high phonon scattering. The scattering of phonons are caused by the impurities, phonons, surface or electrons. The interaction of electron-phonon in semiconductors are negligible. The high phonon scattering also indicates the lower phonon group velocity and reduced thermal conductivity [42-43]. Although, the band dispersion curve of the $Rb_2AuYI_6$ compound has a small negative value for acoustic mode (which also seems to be in the DOS curve), but the mechanical criteria (supplementary section), and the formation energy by element and competing phase strongly suggest the mechanically and thermo-dynamical stability of that compound. So, the studied compounds are considered to be synthesizable as they meets up the following criteria of tolerance and octahedral factor of certain range, negative value of formation and binding energy, and also the negativity of competing phases.

**Table 1**: Structural entity of the compounds under investigation.

| Parameter | $Cs_2AuYCl_6$ | $Cs_2AuYBr_6$ | $Cs_2AuYI_6$ | $Rb_2AuYCl_6$ | $Rb_2AuYBr_6$ | $Rb_2AuYI_6$ |
|---|---|---|---|---|---|---|
| Lattice constant, $a=b=c$ (Å) $a=b=c$ (Bohr) | 10.7811 20.3733 | 11.3529 21.4538 | 12.1495 22.9592 | 10.7261 20.2693 | 11.2605 21.2792 | 12.0988 22.8634 |
| Final volume, V (Bohr$^3$) | 2114.09 | 2468.63 | 3025.59 | 2081.89 | 2408.88 | 2987.88 |
| Tolerance factor, $T_F$ | 0.88 | 0.87 | 0.86 | 0.85 | 0.84 | 0.83 |
| Octahedral factor, $\mu$ | 0.62 | 0.57 | 0.51 | 0.62 | 0.57 | 0.51 |
| Formation energy, $E_F$ (eV/atom) | -2.45 | -2.07 | -1.62 | -2.41 | -2.02 | -1.57 |
| Binding energy, $E_b$ (eV/atom) | -4.18 | -3.81 | -3.41 | -4.16 | -3.78 | -3.37 |
| Minimum energy, $E_0$ (Ry) | -81567.68 | -107309.20 | -161457.95 | -62332.44 | -88073.96 | -142222.70 |
| Bulk modulus, $B_0$ | 32.10 | 27.35 | 22.19 | 33.25 | 27.64 | 22.75 |
| $B'_0$ | 5.14 | 4.60 | 3.12 | 5.49 | 4.36 | 3.17 |

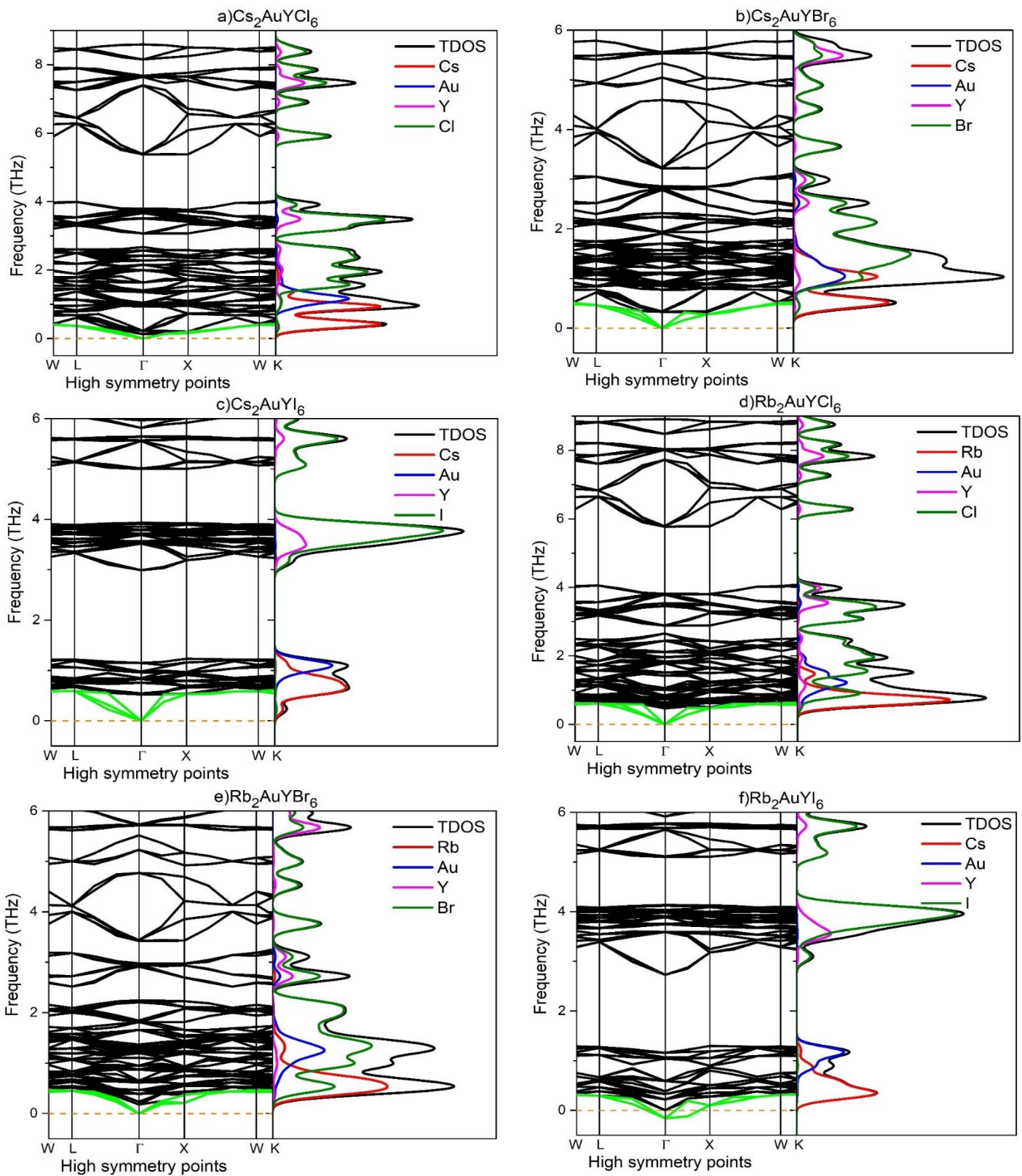

**Fig. 2**: Phonon band dispersion (left side) with TDOS and PDOS (right side) of a) $Cs_2AuYCl_6$ b) $Cs_2AuYBr_6$ and c) $Cs_2AuYI_6$ d) $Rb_2AuYCl_6$ e) $Rb_2AuYBr_6$ f) $Rb_2AuYI_6$

The chemical or thermo-dynamic stability may be calculated by its phase stability. The possible identified competing phases are shown in the right side of equation of Cs and Rb based halides. If any path of formation enthalpy is positive, no stable region is possible in the phase diagram. So, the formation enthalpy by these competing phase and also each of identified phase must be negative. All compounds and phases are shows the negative value, indicating the chemical stability of studied compound (**See Table 2**).

$$\text{For Cs based halides, } Cs_2AuYX_6 \rightarrow \frac{1}{4} CsAuX_3 + \frac{7}{10} Cs_3Y_2X_9 + \frac{1}{20} Au$$

$$\text{For Rb based halides, } Rb_2AuYX_6 \rightarrow \frac{1}{4} RbAuX_3 + \frac{1}{2} Rb_3YX_6 + \frac{1}{5} YX_3 + \frac{1}{20} Au$$

**Table 2**: Formation enthalpy for chemical stability.

| Compound | Phases | Formation enthalpy by competing phase (meV/atom) | Formation enthalpy of each phases (eV/atom) |
| --- | --- | --- | --- |
| $Cs_2AuYCl_6$ | $CsAuCl_3$, $Cs_3Y_2Cl_9$, Au | -25 | -1.67, -2.90, 0 |
| $Cs_2AuYBr_6$ | $CsAuBr_3$, $Cs_3Y_2Br_9$, Au | -18 | -1.43, -2.45, 0 |
| $Cs_2AuYI_6$ | $CsAuI_3$, $Cs_3Y_2I_9$, Au | -12 | -1.16, -1.90, 0 |
| $Rb_2AuYCl_6$ | $RbAuCl_3$, $Rb_3YCl_6$, $YCl_3$, Au | -17 | -1.52, -2.77, -3.20, 0 |
| $Rb_2AuYBr_6$ | $RbAuBr_3$, $Rb_3YBr_6$, $YBr_3$, Au | -10 | -0.86, -2.38, -2.70, 0 |
| $Rb_2AuYI_6$ | $RbAuI_3$, $Rb_3YI_6$, $YI_3$, Au | -5 | -1.03, -1.80, -2.00, 0 |

### 3.2. *Electronic properties, effective mass of charge carrier and density of states*

DFT calculations can provide insights into the band structure, mobility, density of states and other electronic properties of titled materials. Whether a material is an insulator, conductor, or semiconductor depends on how its electrons are distributed throughout their energy levels, which is described by its band structure[44]. The basic need for the realization of optical and thermoelectric characterization is also the understanding of electronic band structure. The distinguishing factor between the valence band (VB) and conduction band (CB) is the Fermi level $E_F$, and all of these indicate the electronic band profile. Determining the band gap, or the energy difference between the highest and lowest unoccupied states, is essential to comprehending the electrical characteristics of the material. The compositions of band profile are displayed in **Fig. 3**. The calculation of band structure was made by four processes: The first is the GGA-PBE SCF calculation, and the second is the TB-mBJ SCF calculation. The band gap values of the GGA-PBE approach are 2.05(2.07), 1.56(1.60), and 1.05(1.07) eV, and those of the TB-mBJ approach are 2.85(2.91), 2.35(2.40), and 1.74(1.78) eV for Cs (Rb) based with halides of Cl, Br and I, respectively. The GGA-PBE functional always shows an underestimated value with correctly structural features, and the TB-mBJ functional is widely recognized and close to experimental value. Thirdly, the band structures computed using TB-mBJ+SOC, which exhibit slightly increased or decreased values (insignificant variation), for the sake of clarification and for convenience (**see Table 3**). These results show a reasonable degree of in accordance with similar studies[29],[45]. Lastly, we also accomplished the band gap values by using HSE-06 functional and the band gap values obtained of 3.03(3.09), 2.48(2.52) and 1.88 (1.92) eV which are more consistent values with TB-mBJ approach (see Fig S2 in supplementary information). So, to determine the exact band gap in present findings, the TB-mBJ potential (mBJ potential) is used, which is more time-efficient than the Hybrid function HSE-06. The outcomes of the HSE-06 potentials exhibit slightly increased band gap values due to the using of different potential, although they are mostly comparable in-line with other findings[46].

Compared to the GGA-PBE strategy, the TB-mBJ and HSE-06 approach for each compound eliminates the degeneracy of certain states, increasing the band gap values. The band gap value from Cl to I, decreases as a result of the overlapping states caused by the increasing size of halogens. It may happen due to the decreasing contribution of Au-*d* and Cl-*p* in the valence band and consequently the gap decrease by the Y-*d* and Au-*s* in the conduction band. Notably, I-based compounds have band gaps that fall within a narrow range (1-2 eV), which makes them useful semiconductors for turning solar energy into electrical power. Additionally, the reported band gap value of $Cs_2TlBiBr_6$ (2.32 eV) is lower than that of $Cs_2TlBiCl_6$ (2.89 eV), which confirming the correctness of our computation[47].

The all compound shows an indirect band gap due to different symmetry point (L-X) by considering the lowest state in CB and highest state in VB for TB-mBJ, TB-mBJ+SOC, and HSE-06 approaches. As a result, the electronic transition from VB to CB occurs possibly by the assistance of phonon (lattice vibration), leading to lower recombination rates[48]. Longer carrier lifetimes may arise from this, which makes them appropriate for certain uses where non-radiative recombination must be kept to a minimum. Although, the presence of phonons is unfavorable since it increases the loss of heat during device operation[49].

The indirect band gap is shown by the electronic band gap threshold energy being smaller than the optical absorption energy. Similar reports were observed in ref.[50]. Besides, the indirect band gap materials may be easier to integrate into current semiconductor devices and circuits since they are more compatible with silicon technology. Additionally, materials with an indirect band gap might have improved thermal stability, which would make them appropriate for high-temperature or harsh environment uses.

**Table 3**: Computed different approaches band gaps, the effective masses of holes ($m_h^*$) and electrons ($m_e^*$), and the effective density of states for studied compounds.

| Compound | Band gap | | | | Effective masses by TB-mBJ approch | | Effective density of states ($\times 10^{18}$ states/cm$^3$) | |
|---|---|---|---|---|---|---|---|---|
| | GGA-PBE | TB-mBJ | TB-mBJ+SOC | HSE-06 | $m_h^*/m_0$ | $m_e^*/m_0$ | $N_v$ | $N_c$ |
| $Cs_2AuYCl_6$ | 2.05 | 2.85 | 2.89 | 3.03 | 0.41 | 0.14 | 6.67 | 1.33 |
| $Cs_2AuYBr_6$ | 1.56 | 2.35 | 2.30 | 2.48 | 0.32 | 0.12 | 4.59 | 1.05 |
| $Cs_2AuYI_6$ | 1.05 | 1.74 | 1.79 | 1.88 | 0.26 | 0.08 | 3.36 | 0.57 |
| $Rb_2AuYCl_6$ | 2.07 | 2.91 | 2.95 | 3.09 | 0.39 | 0.13 | 6.18 | 1.19 |
| $Rb_2AuYBr_6$ | 1.60 | 2.40 | 2.33 | 2.52 | 0.30 | 0.12 | 4.17 | 1.05 |
| $Rb_2AuYI_6$ | 1.07 | 1.78 | 1.81 | 1.92 | 0.25 | 0.07 | 3.18 | 0.47 |

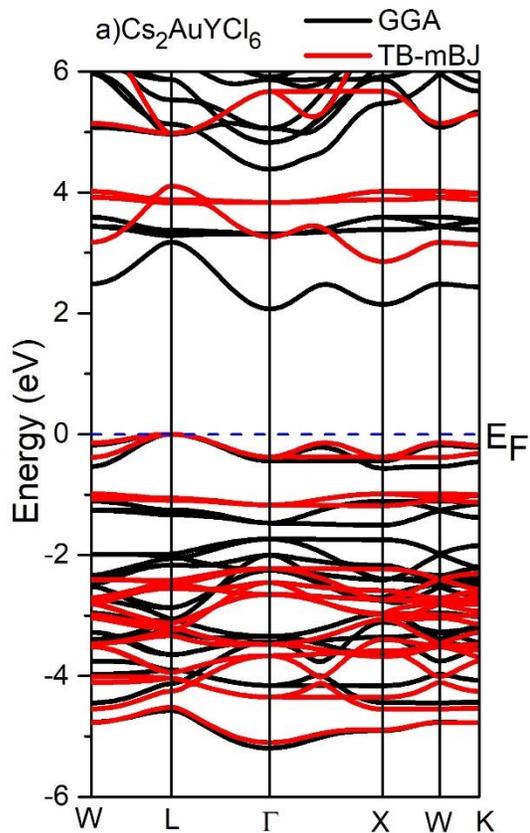
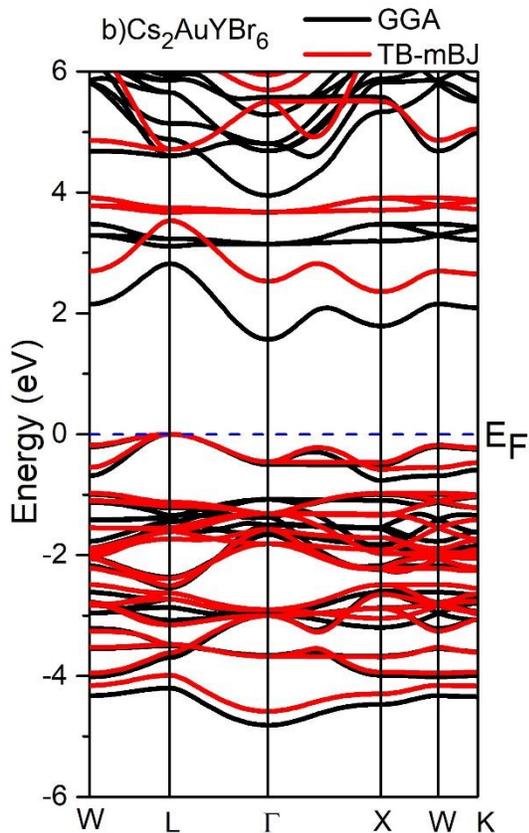
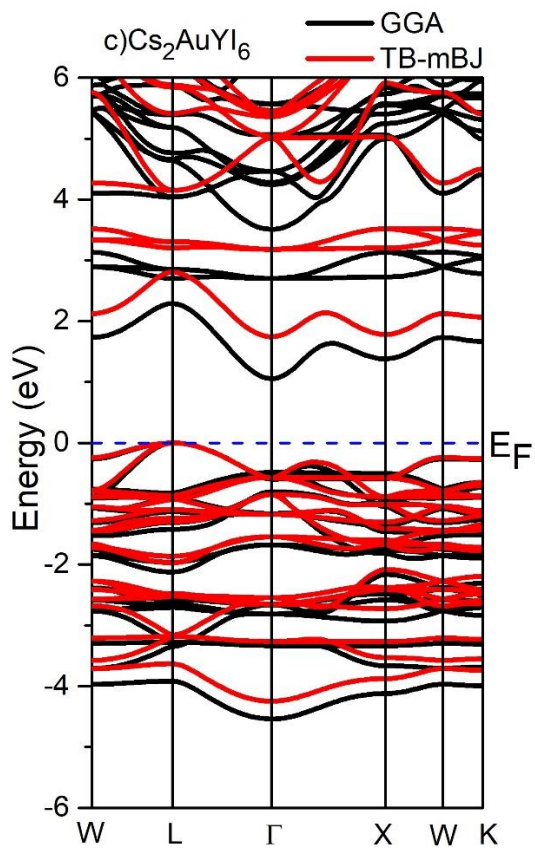
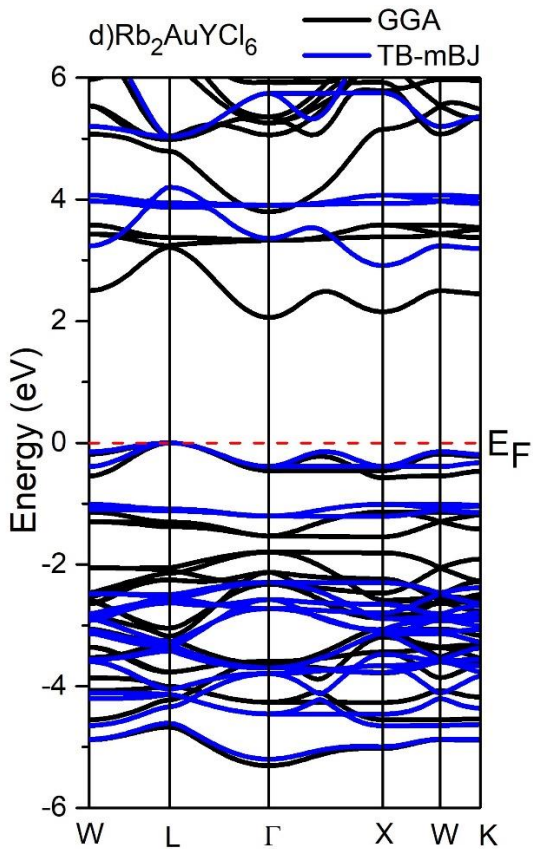

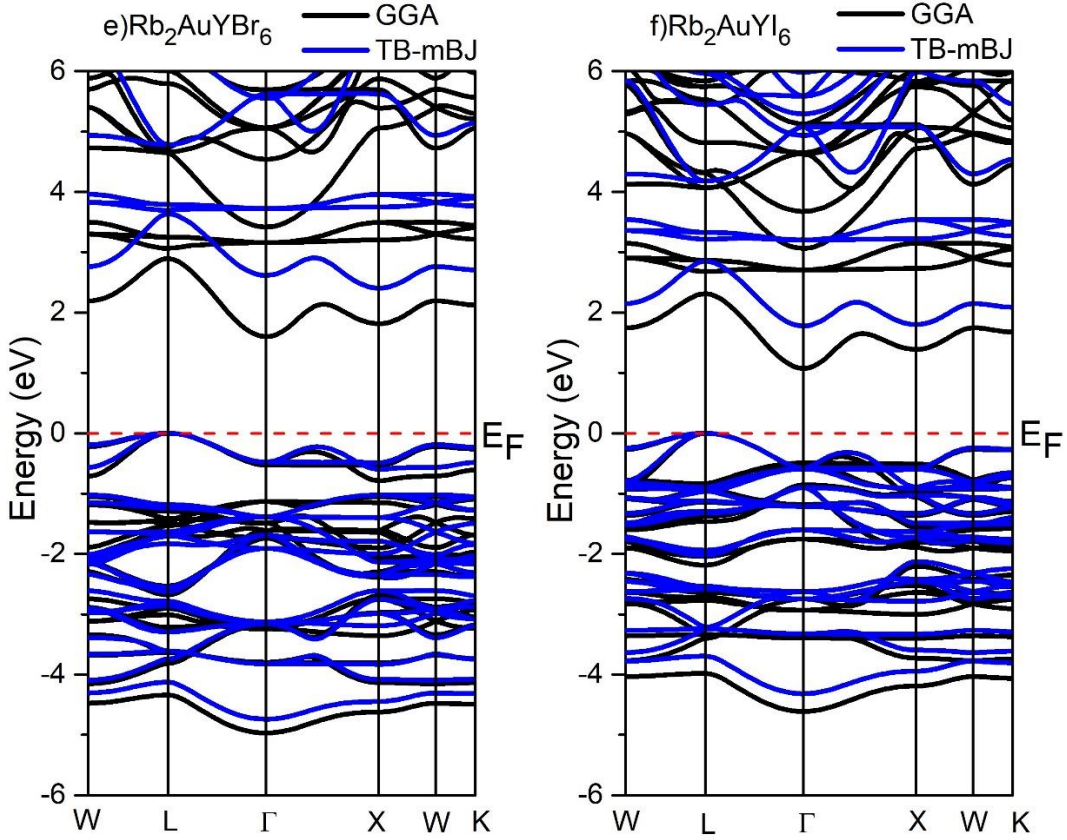

**Fig.3**: Electronic band structure of a) $Cs_2AuYCl_6$, b) $Cs_2AuYBr_6$, c) $Cs_2AuYI_6$ d) $Rb_2AuYCl_6$ e) $Rb_2AuYBr_6$ f) $Rb_2AuYI_6$ computed by GGA-PBE (Black color) and TB-mBJ (Red and blue) approach.

The behavior of charge carriers, like as electrons or holes, is described by their effective mass as if they were mass-specific free particles. The crystal lattice has an impact on charge carriers in crystalline solids, leading to complicated behaviors. The effective mass helps to simplify this complicated situation by treating charge-carriers as if they were flowing in a gas of free electrons or holes with a fixed effective mass. For this, firstly, analyze the band structure to determine the curvature near the band edges. The curvature must be second derivative of the energy (E) with respect to momentum (K), i.e., $d^2E/dK^2$ and the effective mass is then given by the inverse of the curvature[51]:

$$m^* = \frac{\hbar^2}{(d^2E/dK^2)}$$

Where, $\hbar = 1.05 \times 10^{-34}$ J/s. The estimated values of effective mass and holes are also presented in **Table 3**. The effective mass of holes is higher than that of electrons due to planar states in VB and high curvature in CB, which is advantageous for making green energy and optoelectronic devices[52-53]. Furthermore, CB shows a greater dispersion pattern than VB, which improves the materials under study's thermoelectric effectiveness. It is observed that the effective mass values show a significant degree of consistency with other similar findings[29] and the values are low compared with other reported values of $Cs_2AgSbX_6$ (X= Cl, Br, I)[21]. The reduced effective mass confers an advantage in terms of better carrier mobility, which is an exceptionally sought-after characteristic in the advancement of highly effective electrical and optoelectronic systems.

Furthermore, by using the subsequent charge carrier, we may determine the effective state densities of the VB and CB, designated as $N_v$ and $N_c$ individually[54].

$$N_v = 2\frac{(2\pi m_h^* k_B T)^{3/2}}{h^3} \approx 2.5409 \times 10^{19} \left(\frac{m_h^*}{m_0}\right)^{3/2}$$

$$N_c = 2\frac{(2\pi m_e^* k_B T)^{3/2}}{h^3} \approx 2.5409 \times 10^{19} \left(\frac{m_e^*}{m_0}\right)^{3/2}$$

Where T is the temperature in Kelvin, $k_B$ is the Boltzmann value, and h represents the Planck constant. The values of $N_v$ and $N_c$ are shown in **Table 3** and the computed values are very low compared to other similar types of DP materials of $Cs_2TlBiX_6$ (X= Cl, Br and I)[47]. Overall, the perovskite solar-powered devices benefit from low operative state densities in both the conduction and valence bands ($N_v$ and $N_c$) so as to optimize their efficacy.

Total DOS and partial DOS are shown for the formation of band structures in the range of -6 eV to +6 eV, as shown in Fig S3 (supplementary part). Plotting the total DOS against energy allows one to identify characteristics like the band gap and the density of states at the Fermi level. The partial DOS describes the contribution of individual atomic orbitals or elements to the total DOS, and it aids in a more thorough analysis of the bonding and electronic structure. The atomic states were similar in the case of all the studied atom. In the top of the VB, the elemental participation consists of Au-*d* and X-*p* orbitals, whereas the bottom of the CB consists of Au-*s*, Y-*d* and X-*p* orbitals. The $Cs^+$ ions only maintain the structural stability rather than the band structure of the studied compound. So, the electronic insights can guide experimental efforts to tailor these materials for specific applications.

### 3.3. *Nature of bonding*

The compound $X_2AuYZ_6$ (Z = Cl, Br, I) is a member of the double perovskites family. Its bonding nature may be comprehended by analyzing the properties of its component elements and their inter-relationships. The relationship depends on their electronegativity and chemical bonds.

In compounds, the halide ions (Z = Cl, Br, I) have comparatively high electro-negativities, whereas Cs/Rb has a low electronegativity. Usually, an ionic connection occurs between Cs/Rb and Z when Cs/Rb gives electrons to Z, producing Z-anions and $Cs^+/Rb^+$ cations. Due to the high electronegativity difference between cesium and halogens, the X-Z bond is predicted to be mostly ionic. A variety of state of oxidation are possible for both yttrium and gold, and they may exchange electrons to create covalent connections within the structure. These elements' octahedral configuration in the perovskite structure makes electron sharing easier.

The two interpenetrating networks of octahedra bonding between B-Z (Au-Z) and B'-Z (Y-Z), is likely to be covalent. In precisely, the Au-Cl shows the strong covalent and Au-Br and Au-I also shows the less covalence, while the Y-Z shows the covalence with weak ionic bond. Similar literature has been reported in ref [55].

In summary, $X_2AuYZ_6$ exhibits a combination of ionic and covalent bonding. The specific nature of the bonds formed by cesium, gold, yttrium, and the halogen (Z) atoms contributes to the unique properties and stability of this double perovskite structure. The characteristics of the bonding may slightly vary depending on the specific halogen present (Cl, Br, and I), but the general trends mentioned above remain applicable.

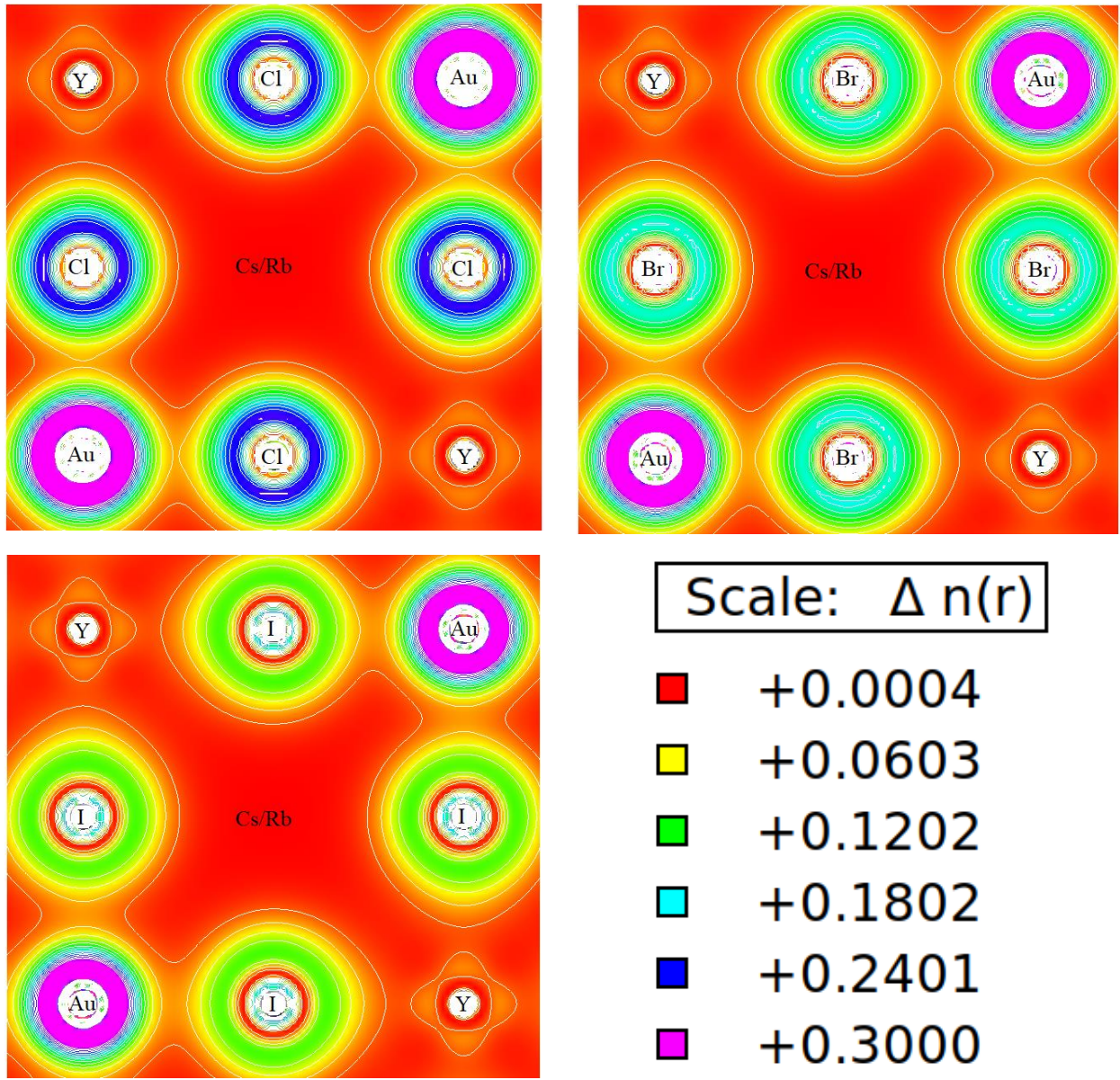

Fig 4: Charge density mapping of Cs/Rb$_2$AuYCl$_6$, Cs/Rb$_2$AuYBr$_6$, and Cs/Rb$_2$AuYI$_6$

### 3.4. *Thermoelectric transport properties*

The ability to efficiently convert waste heat into usable electrical power and vice versa is a crucial feature of the transport or thermoelectric property, which has many real-world applications in a variety of industries, promotes energy sustainability, and helps to solve energy-related problems. In semiconductors, the different transport parameters are determined by band structure, with the band gap from Fermi-level, type of carrier concentration, and carrier effective mass values, making the most significant contributions[56].

To investigate the suitability of compounds for device applications, it is essential to examine their transport properties by considering TB-mBJ Fermi energy. The BoltzTraP2 code incorporates a fixed relaxation time ($\tau$) set at $10^{-14}$ seconds for use in all thermoelectric applications [34]. In the temperature range of 100–1000K, the temperature-dependent variables of electrical conductivity ($\sigma/\tau$), carrier concentration (N), and electronic conductivity, ($k_e/\tau$), are calculated and presented in **Fig. 5(a–c)**.

The electrical conductivity which characterizes the ability of the material to conduct electricity. A rise in carrier concentration (N) causes an upward trend in the electrical conductivity as a function of temperature and may be expressed as: $\sigma = Ne\mu$, which is characteristic of semiconducting material. Moreover, when carrier concentration is positive, it implies that holes are contributing, and when it is negative, it shows that electrons are the main carrier[57]. As the temperature rises, the electrons acquire energy, resulting in the strong positive variation of carrier concentration. The reported $\sigma/\tau$ values at 300K are 0.94(1.01) $\times 10^{19}$, 1.02(1.02) $\times 10^{19}$ and 0.92(0.90) $\times 10^{19}$ $(\Omega.m.s)^{-1}$ for Cs and Rb based with halides (Cl, Br, I). The study has shown that, when Cl is replaced with Br and I, the value of $\sigma$ increased with temperature. But in the case of I of both compounds, the value of $\sigma$ crosses at a certain level (350K for Cl and 450K for Br) and shows the higher value with high temperature. The aforementioned compounds exhibit comparatively high electrical conductivity as a result of their incredibly low resistances; this suggests that they may find application in the fabrication of thermoelectric systems.

The Boltztrap2 code solely focuses on the electronic part of thermal conductivity ($k_e/\tau$), as phonon calculations go beyond the scope. The calculated values of $k_e/\tau$ at 300K for $Cs_2AuYCl_6$ ($Rb_2AuYCl_6$), $Cs_2AuYBr_6$ ($Rb_2AuYBr_6$), and $Cs_2AuYI_6$ ($Rb_2AuYI_6$) are 0.36 (0.40) $\times 10^{14}$, 0.42 (0.42) $\times 10^{14}$ and 0.39 (0.38) $\times 10^{14}$ ($Wm^{-1}k^{-1}s^{-1}$), respectively. As the temperature of a material rises, the lattice vibrates progressively more, which raises the electronic component of thermal conductivity. It should be noted that a factor of $10^5$ (the ratio of $\sigma$ and $k_e$) indicates that the materials are capable of producing thermoelectric power. The $\sigma$ and $k_e$ leads to the identical pattern as the both are rely on carrier concentration (N).

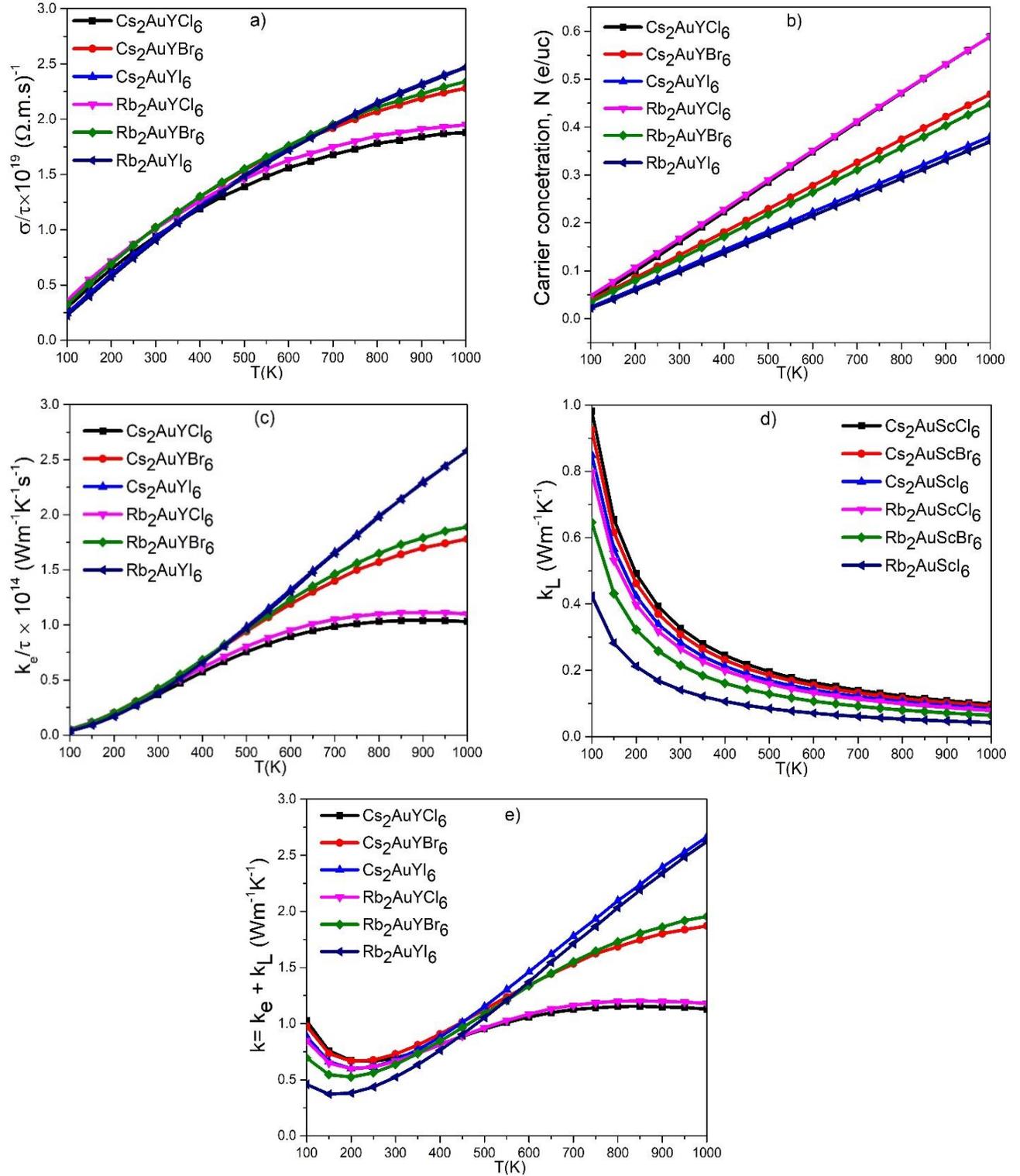

**Fig.5**: The computed a) **σ/τ**: Electrical conductivity, b) **N**: Carrier concentration, c) **k_e/τ**: electronic thermal conductivity, and d) **k_L**: Lattice thermal conductivity and e) **k**: Total thermal conductivity, of the studied compounds.

The total thermal conductivity (k) of a material may be calculated by considering the lattice thermal conductivity ($k_L$) and electronic thermal conductivity ($k_e$), where $k = k_e + k_L$. Slack's model is very popular and widely used to evaluate the $k_L$ independently and can be estimated using the following formula [58]:

$$k_L = \frac{A(\gamma)\delta M_{av}\theta_D^3}{\gamma^2 n^{2/3} T}$$

A is a constant dependent on $\gamma$, $\delta$ is the mean atomic volume of cubic root, $M_{av}$ is the mean atomic mass, $\theta_D$ is the Debye hardness, $\gamma$ corresponds to the Gruneisen parameter, n stands for the unit cell atom and T represents the absolute temperature in kelvin. **Figure 5 (d)** plots the computed value of $K_L$ versus T and demonstrates how it decreases throughout the whole temperature range. Increasing temperature weakens interatomic lattice links, leading to a decrease in $K_L$ and phonon group velocity. Besides, there is an inverse relationship between temperature and lattice thermal conductivity ($k_L \propto 1/T$)[59-60]. The corresponding values of $k_L$ at room temperature for Cs (Rb) based halides are 0.32 (0.26), 0.30 (0.21), and 0.28 (0.14) $Wm^{-1}k^{-1}$, respectively.

The overall thermal conductivity of the material under investigation is illustrated in **Fig 5 (e)**. It first declines at low temperatures before significantly increasing between 200 and 1000K in relation to electrical conductivity. Because of this, the Wide-man Franz law supports these conclusions to the charge carrier or electronic conductivity and expresses their proportionate relationship as follows: $k_e = LT\sigma$ [61], where L is the Lorentz constant ($2.44 \times 10^{-8}$ W/S-$K^2$). However, the Rb-based compound exhibits lower thermal conductivity than Cs-based halides at ambient temperature (300K), which might have an impact on the material's efficiency.

The see-beck coefficient (S) of degenerative semiconducting material is measured by the following equation, which takes into account the combined impact of carrier concentration (N) and the effective mass (m)[62]:

$$S = \left(\frac{8\pi^2 k_B^2}{3h^2 e}\right)\left(\frac{3N}{\pi}\right)^{-2/3} m^* T$$

In general, materials with a smaller band gap show greater S values, whereas those with a larger band gap show lower S values[63-64]. It is clear that, the value of S decreases, as the S and N are inversely connected, due to the rising of intrinsic carrier, as shown in **Fig 6(a)**. The values of S at room temperature for $Cs_2AuYCl_6$ ($Rb_2AuYCl_6$), $Cs_2AuYBr_6$ ($Rb_2AuYBr_6$), and $Cs_2AuYI_6$ ($Rb_2AuYI_6$) are 195(189), 195 (197), and 206(209) µV/K, respectively. The PDOS suggests that these substances are p-type semiconductors based on their positive S values which can be used for power generation applications.

The power factor is a significant factor for TE compounds and systems. The calculation is commonly represented by the symbol "PF" and is as follows: power factor (PF) = ($S^2\sigma$). An efficient thermoelectric material for power-generating applications is indicated by a greater power factor. High electrical conductivity and a high See-beck coefficient are desirable qualities for thermoelectric materials in order to maximize power factor. It may be difficult to optimize both features, however, since their relationships are often inverse. The value of PF recorded at 300K are 3.57 (3.62) ×$10^{-3}$, 3.88(3.94) ×$10^{-3}$ and 3.94 (3.95) ×$10^{-3}$ ($Wm^{-1}k^{-2}$), respectively, as presented in **Fig 6(b)**.

A critical indicator for assessing how well a thermoelectric material convert's heat into electricity or efficiency is the figure of merit (ZT). The formula for the figure of merit is, $ZT= S^2\sigma T/(k_e+k_L)$[65]. This dimensionless value considers the See-beck coefficient (S), thermal conductivity (k), and electrical conductivity ($\sigma$), all of which are essential properties of a thermoelectric material. Scientists and engineers strive to find a balance that maximizes the ZT value, however optimizing these features may be difficult since they often have contradictory correlations. A thermoelectric material with a higher ZT value (~1) is said to be more efficient for use in thermoelectric applications, including waste heat generators or thermoelectric cooling systems for refrigeration purposes.

By utilizing the electronic and lattice contributions, we have calculated the actual ZT value. Consequently, the dimensionless ZT value falls as a consequence of an increase in total thermal conductivity, k. The actual greatest ZT values are found to be 0.51 , 0.53, and 0.58 for Cs based halides (Cl, Br, I) and 0.55, 0.62 and 0.75 for Rb based halides (Cl, Br, I) at 300K, respectively which are higher than the metal halide perovskite materials of $CsSnI_3$ (ZT = 0.14)[66], $FASnI_3$ (ZT= 0.19)[67], and $CsSn_{0.8}Ge_{0.2}I_3$ (ZT = 0.12)[68], as shown in **Fig. 6(c)**. Additionally, these outcomes are in line with comparable categories of published resources such as $Cs_2AuScX_6$ (X= Cl, Br, I; ZT = 0.44, 0.47, 0.62)[29], $Rb_2AuScX_6$ (X= Cl, Br, I; ZT = 0.47, 0.53, 0.54)[29], and $Rb_2LiTlX_6$ (X = Cl. Br; ZT= 0.43/0.51)[69] at room temperature, and also carry greater value than other DPs halide materials such as $Rb_2AuBiX_6$ [70], $Li_2CuBiZ_6$ [71], $Cs/Rb_2KScI_6$ [72], $Rb_2ScTlX_6$ [24], and $Cs_2InAsX_6$ [73]. It should be noted that the theoretical calculation of $k_L$ using Slacks equation overestimates the actual value largely[46], but in most cases such as SnSe [74], and $Ag_2XYSe_4$ [46], the BTE code and experimental works have negligible difference for the calculation of lattice thermal conductivity. Therefore, it is our expectation that the experimental ZT values will exceed the values that we computed.

When a thermoelectric material is subjected to temperature variations, its specific heat affects its capacity to absorb or release heat. The specific heat of an ideal thermoelectric material should be as low as feasible. This is due to the fact that a material with a reduced specific heat becomes more responsive to changes in temperature as it requires less thermal energy to increase its temperature. The maximum value of Cv is shown for I-based composition, and it rises steadily as T increases. The majority of solids attain their saturation value as the temperature rises above the Debye temperature; however, highest saturation was seen for I-based compounds than Br and Cl based compounds in this range of temperatures, as shown in **Fig 6 (d)**.

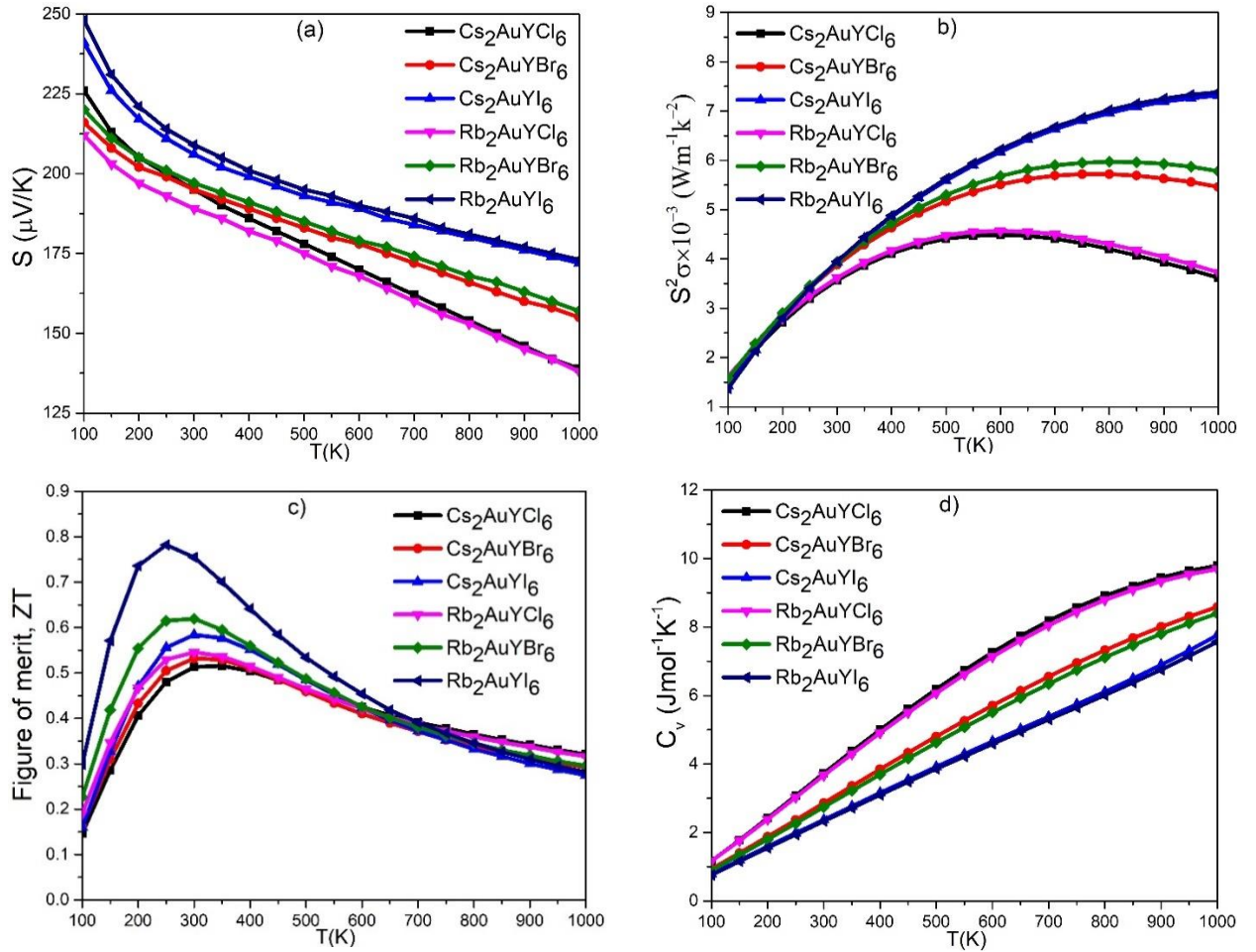

**Fig.6**: a) **S**: See-beck coefficient, b) **S²σ/τ**: Power factor, c) **ZT**: Figure of merit, d) **Cv**: Constant-volume specific heat, of the studied compounds.

### 3.5. *Optical properties*

Any materials response to light is described by its optical properties which directly influence the device performance. The optimized optical nature may be identified by the complex quantity of dielectric tensor, which is given by the Ehrenreich and Cohen's equation [75]: $\epsilon(\omega) = \varepsilon_1(\omega) + i\varepsilon_2(\omega)$, where ω represents the angular frequency, $\varepsilon_1(\omega)$ and $\varepsilon_2(\omega)$ represents the real and imaginary component of dielectric tensor, calculated by the Kramers-Kronig [76] and Kohn-Shams equations[77], which are responsible for the polarization and light absorption.

The static real dielectric constant $[\varepsilon_1(0)]$ was found to be 2.44(2.36), 3.01(2.96) and 4.01(3.95) which confirms the highest dispersion at resonance frequency as shown in **Fig. 7(a)**. The enhanced reactivity of a substance towards incident energy is attributed to the greater value of $\varepsilon_1(0)$. The examined substance is of high importance in the visible and ultraviolet range, as shown by the peak value of $\varepsilon_1(\omega)$. When the materials go from Cl to Br to I, there is a greater electronic impact, which causes the peak value (in eV) to fall. In addition, the Penn's model links the band gap with static polarization in an inverse connection[78].

$$\varepsilon_1(0) \approx 1 + \left[\frac{\hbar\omega_p}{E_g}\right]^2$$

The $\varepsilon_2(\omega)$ is presented in **Fig 7(b)**. It governs the amount of absorbed radiation, and its peak values convey the shift between the valence band maxima and the conduction band minima, as well as the losses of energy within these compounds. The optical band gap is shown by a threshold energy value calculated by $\varepsilon_2(\omega)$ for the compounds under study, which is 2.92(2.98), 2.38(2.43), and 1.75(1.80) eV. This value seems to be very near to the estimated electronic band gap. A slight overestimation of the optical band gap occurs in optical properties as a result of the calculation of its related potential. In the present findings, it shows a highly significant degree of accuracy.

Utilizing $\varepsilon_1(\omega)$ and $\varepsilon_2(\omega)$, one may compute other optical parameters such as complex refractive index, absorption coefficient $\alpha(\omega)$, real optical conductivity $\sigma(\omega)$, optical reflectivity R(ω), and electron energy loss L(ω). The formula for the complex refractive index is n(ω)-ik(ω), where k(ω) is the imaginary component (extinction coefficient) and n(ω) is the real part (refractive index). The following equations calculated the optical constants using the conventional symbol meanings[53],[79-80]:

$$n(\omega) = \frac{1}{\sqrt{2}} \left\{ \sqrt{\varepsilon_1^2(\omega) + \varepsilon_2^2(\omega)} + \varepsilon_1(\omega) \right\}^{\frac{1}{2}}$$

$$k(\omega) = \frac{1}{\sqrt{2}} \left\{ \sqrt{\varepsilon_1^2(\omega) + \varepsilon_2^2(\omega)} - \varepsilon_1(\omega) \right\}^{\frac{1}{2}}$$

$$\alpha(\omega) = \sqrt{2}\omega \left\{ \sqrt{\varepsilon_1^2(\omega) + \varepsilon_2^2(\omega)} - \varepsilon_1(\omega) \right\}^{\frac{1}{2}}$$

$$\sigma(\omega) = \frac{\alpha(\omega) * n(\omega) * c}{4\pi} \approx \frac{\varepsilon_2(\omega)}{4\pi}$$

$$R(\omega) = \left| \frac{\sqrt{\varepsilon_1(\omega) + i\varepsilon_2(\omega)} - 1}{\sqrt{\varepsilon_1(\omega) + i\varepsilon_2(\omega)} + 1} \right|^2$$

$$L(\omega) = \frac{\varepsilon_2(\omega)}{\varepsilon_2^2(\omega) + \varepsilon_1^2(\omega)}$$

The real refractive index is determining how much light is bent or refracted as it passes from one medium to another i.e. measures the transparency of material. The variation of energy with n(ω) are presented in **Fig 7(c)**. The n(ω) and $\varepsilon_1(\omega)$ shows the similar behavior in the following link[81]: $n^2(0) = \varepsilon_1(0)$. The refractive index has static values of 1.55(1.52), 1.74(1.71), and 2.00(1.97) and maximum values of 1.81(1.78), 2.06(2.03), and 2.45(2.41) in the first absorption peak, which are more suited in comparison with silicon nitride (1.9) for solar cell applications[82]. Additionally, the transparency (i.e lowest reflectivity and covalence nature) was also exists due to the greater value of 1 of n(ω).

The extinction coefficient quantifies the extent to which a material absorbs and scatters light. It is associated with absorption processes, and a higher k value indicates stronger absorption. As shown in **Fig 7(d)**, the first highest extinction coefficient is seen in visible to UV spectrum. It should be noticed that the greatest value of k(ω) is located approximately at the zero value of $\varepsilon_1(\omega)$.

The absorption coefficient, α(ω), in semiconductors is important to comprehend how materials take in photons and produce electron-hole pairs. Devices like solar cells and photodiodes depend on this mechanism to function. The amount of incoming light that a substance absorbs is measured by its absorption coefficient. Stronger absorption is indicated by a greater absorption coefficient, which results in a more notable decrease in the transmitted light intensity. Higher band gap materials are known to exhibit reduced absorption in the visible range of electromagnetic waves[83]; this is also present with the compound under study. In the visible photon energy range of 1.77 to 3.26 eV with corresponding wavelength of 700 to 380 nm, the compounds of Cl, Br and I-based of Cs(Rb) exhibit maximum absorption coefficients of 0.23(0.35) ×$10^5$ $cm^{-1}$, 1.01(1.06) ×$10^5$ $cm^{-1}$, and 1.42(1.48) × $10^5$ $cm^{-1}$, respectively, as shown in **Fig 8(a) and 8(b)**. These results are very consistent with solar cell materials (Si, GaAs, InP, CdTe)[84] and others DP halide absorbing materials of $Rb_2LiTlX_6$[69], $Cs_2MGaBr_6$[85], and $Cs_2KInX_6$[86]. Moreover, the all compounds also exhibits a different increasing absorption coefficient in the UV spectrum of 4 to 12 eV and the I- based compound always shows the maximum absorption peak as well. When an electron moves from its valence band (VB) to its conduction band (CB), a material with a greater absorption coefficient reacts more positively.

**Table 4**: Calculated optical constants

| Compound | $\varepsilon_1(0)$ | n(0) | R(0) |
|---|---|---|---|
| $Cs_2AuYCl_6$ | 2.44 | 1.55 | 11.14% |
| $Cs_2AuYBr_6$ | 3.01 | 1.74 | 7.15% |
| $Cs_2AuYI_6$ | 4.01 | 2.00 | 4.69% |
| $Rb_2AuYCl_6$ | 2.36 | 1.52 | 10.86% |
| $Rb_2AuYBr_6$ | 2.96 | 1.71 | 6.97% |
| $Rb_2AuYI_6$ | 3.95 | 1.97 | 4.41% |

The primary source of optical conductivity in semiconductors is inter-band shifts, which occur when photons are absorbed and excites electrons from the VB to the CB due to breaking bond. Different optical conductivity peaks are observed in the photon energy range of 0 to 12eV while the largest peak are at the energy of 6-8 eV and 7-11 eV for Cl, 5-6 eV and 8-10 eV for Br, and 4-5 eV and 7-11 eV for I-based compounds, as presented in **Fig 8(c)**. These peaks are quite interesting and may have applications in optoelectronics since they exist at different energy levels. The fact that both conductivity and absorption possess minimum and maximum in comparable regions validates the concept of theory and, therefore, the precision of the calculated outcomes.

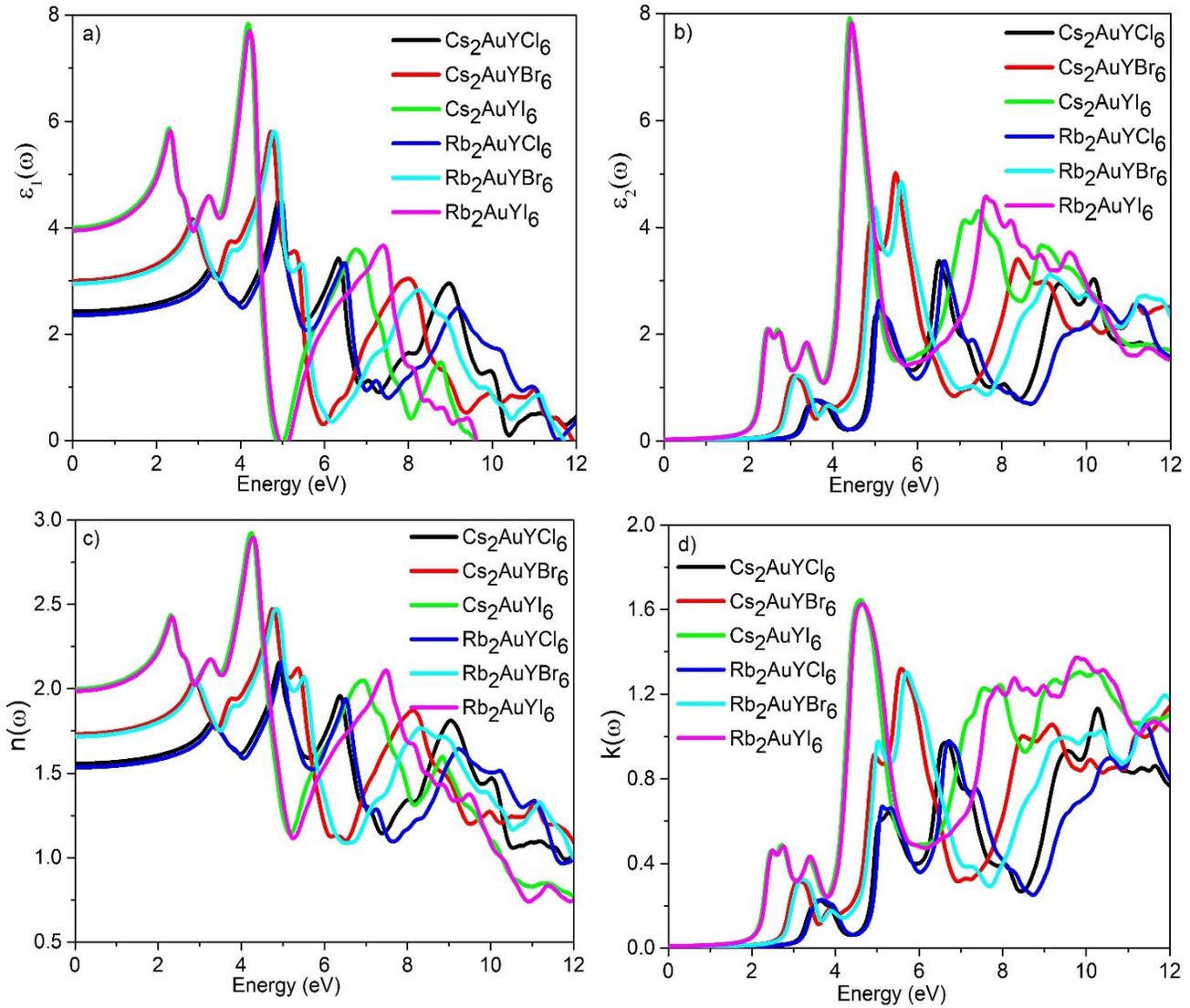

**Fig.7**: a) $\varepsilon_1(\omega)$: real part of the dielectric function, b) $\varepsilon_2(\omega)$: imaginary part of the dielectric function, c) $n(\omega)$: real part of refractive index, and d) $k(\omega)$: imaginary part of refractive index with photon energy for all studied compounds.

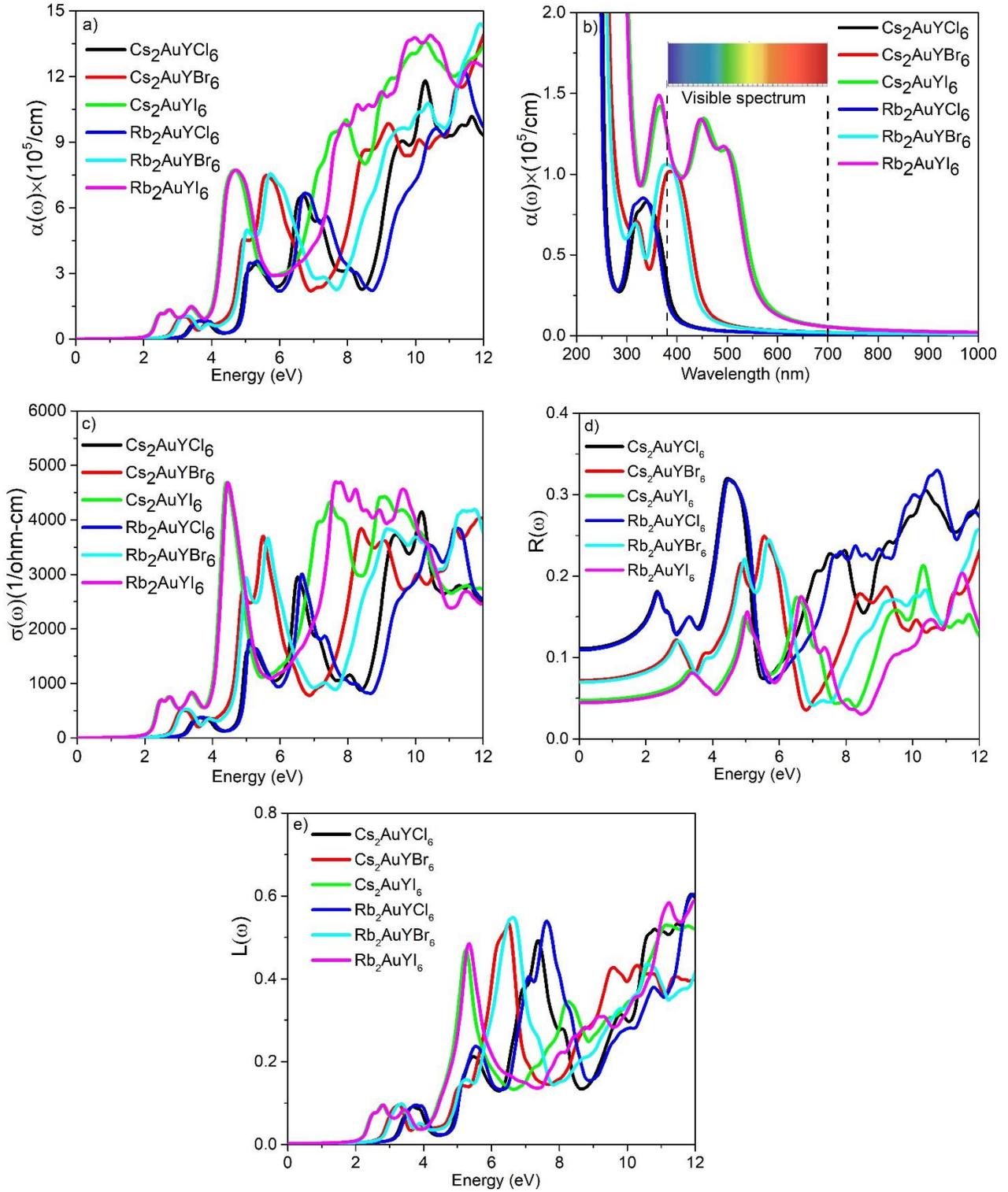

**Fig. 8**: a) and b) **α(ω)**: Absorption coefficient w.r.t. eV and nm, c) **σ(ω)**: Real optical conductivity, d) **R(ω)**: Optical reflectivity, and e) **L(ω)**: Electron energy loss of studied compounds.

The quantity of incoming light that is reflected at a semiconductor material's surface is known as optical reflectivity, R(ω). **Fig 8(d)** shows the optical reflectivity with incident energy variation and shows the inverse connection of absorbance. It demonstrates that the cut-off reflectivity's values are 0.11(0.10), 0.071(0.069), and 0.046 (0.044) for Cs (Rb) based increasing halide. The reflectivity expanded with increasing photon energy (eV), with the named compounds (both series of Cl, Br and I) showing maximum reflectivity of 18%, 12% and 8.5% in the visible range (1.75-3.10 eV). For comparison, the Si solar cell shows 20–13.9% reflectance with respect to 400–700 nm, which implies the more prominent nature of our studied compound[87].

The energy loss of moving electrons within the material is seen in **Fig. 8(e)**. It is observed that there is no static loss and that the losses begin at the corresponding threshold value, reaching the maximum value of 0.09 in the range of 0 to 4eV for all compounds. Additionally, the losses increase with increasing energy and it is high (50%) above 5 eV.

**3.6** *Spectroscopic limited maximum efficiency (SLME)*: The photovoltaic theoretical efficiency of any absorbing materials are calculated by utilizing the SLME model. A straightforward concept for measuring efficiency via band gap is provided by the S.Q limit[88]. The extended S.Q model of indirect or direct band gap materials with coefficient of absorption spectra, film thickness, and radiative or non-radiative losses is taken into account by the SLME model[89-90]. The following equations are employed for theoretical power conversion efficiency:

$$\eta = \frac{P_m}{P_{in}} = \frac{Maximum\ output\ power}{Total\ incident\ power}$$

In SLME calculation, the maximum output power ($P_m$) is determined by taking into account the I-V curve of the solar cell[91].

$$P_m = J_{sc} \times V_{oc} \times FF$$

Where, *Jsc*, *Voc* and *FF* stands for short circuit current, open circuit voltage and fill factor. The following expression are used for short circuit current, *Jsc* [92]and and the total incident power, Pin, assumed to be 1000 Wm$^{-2}$[93]:

$$J_{sc} = e \int_{Eg}^{\alpha} \alpha(E) I_{sun}(E) dE$$

Where e is the charge, $\alpha(E)$ is the photon absorptivity and Isun is the current density of AM1.5G solar spectrum at room temperature.

The efficiency impact on these materials as a function of absorbent layer thickness is calculated using the aforesaid method. For the materials under study, the thickness variation ranges from 0 to 1.4μm, as shown in **Fig 9**. The findings demonstrate that the compounds based on Cl exhibit the maximum efficiency of 8.87% and 8.69% for $Cs_2AuYCl_6$ and $Rb_2AuYCl_6$, while the compounds based on Br exhibit 12.04% and 11.58 at 0.4μm, respectively, for $Cs_2AuYBr_6$ and $Rb_2AuYBr_6$. It is worth noting that the I-based double perovskite achieves a better power conversion efficiency of around 25.20% and 24.90% for $Cs_2AuYI_6$ and $Rb_2AuYI_6$ at 1μm thickness, respectively, due to its small band gap and higher absorption coefficient than Br and Cl-based compounds.

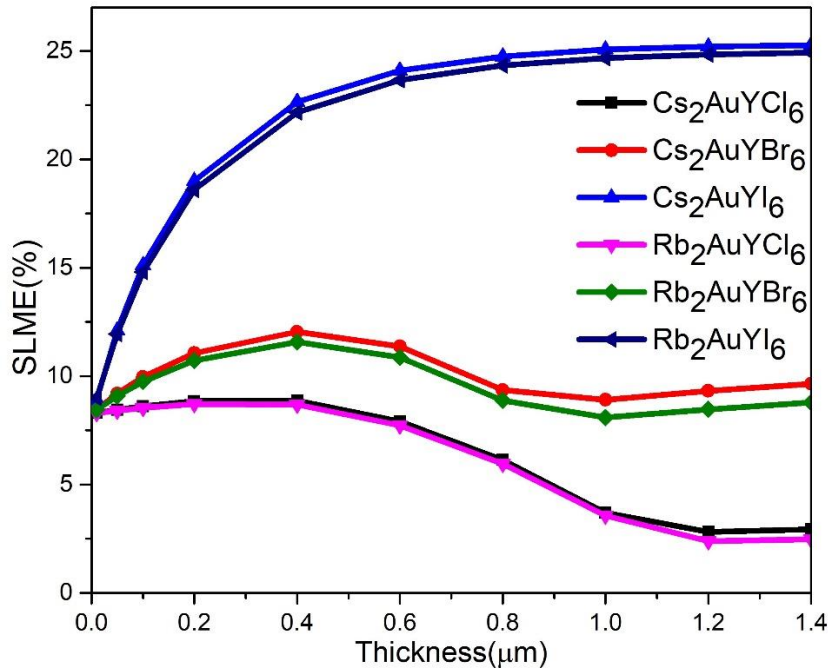

**Fig. 9**: Efficiency variation with absorbing layer thickness.

## 4. *Conclusion*

DFT analysis was used to look into the optimized structure of $X_2AuYZ_6$ (X = Cs/ Rb; Z= Cl/Br/I) for opto-electronic and thermoelectric power generators. The structural integrity of the cubic phase stability is confirmed by many criteria, including tolerance factors ranging from 0.83 to 0.88 and octahedral factors varying from 0.51 to 0.62. Their positive phonon dispersion, negative formation and binding energy were used for assessing the thermodynamic and dynamic equilibrium. The ductility is verified by the mechanical behavior, as both the B/G ratio (>1.75) and σ (>0.26) surpass the critical limit. The decrease in band gap [2.82(2.91), 2.35 (2.40) for Cs (Rb) based halides was validated by tuning the indirect band gap with the replacement of halogens (Cl, Br and I) in the L-X symmetry points. In CBM, the orbital contribution mainly comes from the Y-d, Au-s, and X-p while the Au-d contributes to the VBM of the band structure. The lower effective charge carrier and state densities obtained signifies the fabrication of clean energy technology. The presence of a negative dielectric constant in the UV range of photon energy allows for its use in optical systems such as filters, fiber optics, and shielding for electromagnetic devices. The refractive index at zero energy was evaluated to be 1.55 (1.52), 1.74(1.71), and 2.00(1.97) for $Cs_2AuYCl_6$ ($Rb_2AuYCl_6$), $Cs_2AuYBr_6$ ($Rb_2AuYBr_6$), and $Cs_2AuYI_6$ ($Rb_2AuYI_6$), respectively. These materials exhibit outstanding absorption values in the visible spectrum with $10^5$ orders and may therefore be employed for opto-electronic applications. The I-based of both compounds (Cs and Rb) shows the higher conversion efficiency of 25.2% and 24.9% for photovoltaic purposes than Br and Cl based compounds, which is based on SLME model. The high figure of merit (ZT) came from a careful study and determination of the transport properties, which included the high-power factor, positive big See-beck coefficient, significant low thermal conductivity, and high electrical conductivity. Comparatively, the ZT values of the Rb-based compounds with same halides (Cl, Br and I) were found to be higher than those of Cs, with values of 0.55 (0.51), 0.62 (0.53), and 0.75 (0.55) recorded at temperatures of 300 K, accordingly. Finally, the narrower band gap compounds, such as $X_2AuYI_6$, makes them suitable for efficient absorption of sunlight and subsequent electricity generation, which also supports for wasted heat management system. Therefore, the studies demonstrate the significance of these double perovskite halides for a broad range of applications in the fields of optoelectronics and thermo-electrics, which will allow for the development of devices for the harvesting of clean or green energy in the future.